\documentclass[conference]{IEEEtran}
\IEEEoverridecommandlockouts
\usepackage{cite}
\usepackage{amsmath,amssymb,amsfonts}
\usepackage{algorithmic}
\usepackage{graphicx}
\usepackage{textcomp}
\usepackage{threeparttable}
\usepackage{xcolor}
\def\BibTeX{{\rm B\kern-.05em{\sc i\kern-.025em b}\kern-.08em
    T\kern-.1667em\lower.7ex\hbox{E}\kern-.125emX}}

\begin{document}

\title{Data-Driven Graph Switching for Cyber-Resilient Control in Microgrids}

\author{\IEEEauthorblockN{Suman Rath}
\IEEEauthorblockA{\textit{Department of Computer Science and Engineering} \\
\textit{University of Nevada, Reno}\\
Reno, NV 89557, USA \\
E-mail: sumanr@unr.edu}
\and
\IEEEauthorblockN{Subham Sahoo}
\IEEEauthorblockA{\textit{Department of Energy}\\
\textit{Aalborg University}\\
Aalborg, 9220, Denmark \\
E-mail: sssa@energy.aau.dk}
}

\maketitle

\begin{abstract}
\textcolor{black}{Distributed microgrids are conventionally dependent on communication networks to achieve secondary control objectives. This dependence makes them vulnerable to stealth data integrity attacks (DIAs) where adversaries may perform manipulations via infected transmitters and repeaters to jeopardize stability.}
This paper presents a physics-guided, supervised Artificial Neural Network (ANN)-based framework that identifies communication-level cyberattacks in microgrids by analyzing whether 
incoming measurements will cause abnormal behavior of the secondary control layer. 
If abnormalities are detected, 
an iteration through possible spanning tree graph topologies that can be used to fulfill secondary control objectives is done. Then, a communication network topology that would not create secondary control abnormalities is identified and enforced for maximum stability.
\textcolor{black}{By altering the communication graph topology, the framework eliminates the secondary control layer’s dependence on inputs from compromised cyber devices helping it achieve resilience without instability. 
Several case studies are provided showcasing the framework's robustness against False Data Injections and repeater-level Man-in-the-Middle attacks.}
To understand practical feasibility, robustness is also verified against larger microgrid sizes and in the presence of varying noise levels.
Our findings indicate that performance can be affected when attempting scalability in the presence of noise. However, the framework operates robustly in low-noise settings.

\end{abstract}


\section{Introduction}

\textcolor{black}{Microgrids are cyber-physical systems with a hierarchical control framework that involves primary and secondary layers for voltage/frequency control and power-sharing regulations \cite{teng2024distributed}. The microgrid secondary control layer is responsible for set-point tracking and relies on communication devices for nominal operations \cite{rocabert2012control}. This makes the system vulnerable to stealth attacks compromising communication devices and manipulating data flow patterns} \cite{tan2016survey}. Under the attacks' influence, this layer computes erroneous control signals that propagate further to jeopardize nominal operation \cite{sahoo2019cyber}.
\textcolor{black}{A necessary requirement for convergence of secondary control inputs is to ensure a spanning tree in the cyber (communication graph) topology \cite{zhang2015optimal}. If this spanning tree relies on compromised network devices, then it would feed untrustworthy inputs to the secondary controller, forcing it to compute erroneous control signals \cite{selfhealing}.}
Hence, it is essential to ensure that the communication graph topology on which the microgrid secondary control layer is dependent is free from manipulations in the cyber layer \cite{kirti1,kirti2}. 

To achieve the objective, this paper presents a physics-guided Artificial Neural Network (ANN) framework that can identify the trustworthiness of the default communication topology by estimating abnormal secondary control outputs that it might create within the microgrid network. {In this context, physics-guided means that the rationale behind using the ANN is rooted in the principles of (domain-specific) microgrid control dynamics. The microgrid local parameters are normally synchronous in the steady state as this is an essential objective of cooperative control action. However, \cite{danzi2016impact} has already established that DIAs and other attack vectors like jamming lead to the disruption of cooperative synchronization \cite{sahoo2018stealth}. This may be reflected as high error outputs from local secondary controllers.} Hence, we seek to estimate the total sum of these outputs (via ANN-assisted regression) before the attack propagates to the secondary control layer. If the total sum is estimated to be unconventionally higher than the expected value (where the expected value is determined from microgrid steady-state behaviors during normal operation), a trigger is generated indicating the possible presence of a cyberattack. On the generation of a trigger, the proposed ANN model iterates through possible spanning tree graph topologies (each of which relies on a distinct set of network devices) to identify a topology that can achieve nominal functionality in a trustworthy manner. This topology is then enforced in the microgrid environment isolating and mitigating the cyberattack.
We provide several case studies highlighting the proposed method's resilience against False Data Injection (FDI) \cite{chlela2016real} and Man-in-the-Middle (MITM) attacks \cite{conti2016survey}.
We also analyze the performance of the proposed framework when scaled up to larger microgrid sizes and in the presence of varying levels of noise.






\section{Control Structure and Attack Formulation}
As shown in Fig. \ref{CONTROL}, this paper considers a conventional two-layered hierarchical control structure consisting of primary and secondary layers. A detailed description of their operational principles and functionalities is provided below:

\begin{figure}[tp]
\centering
    \centering
    \includegraphics[width=\linewidth]{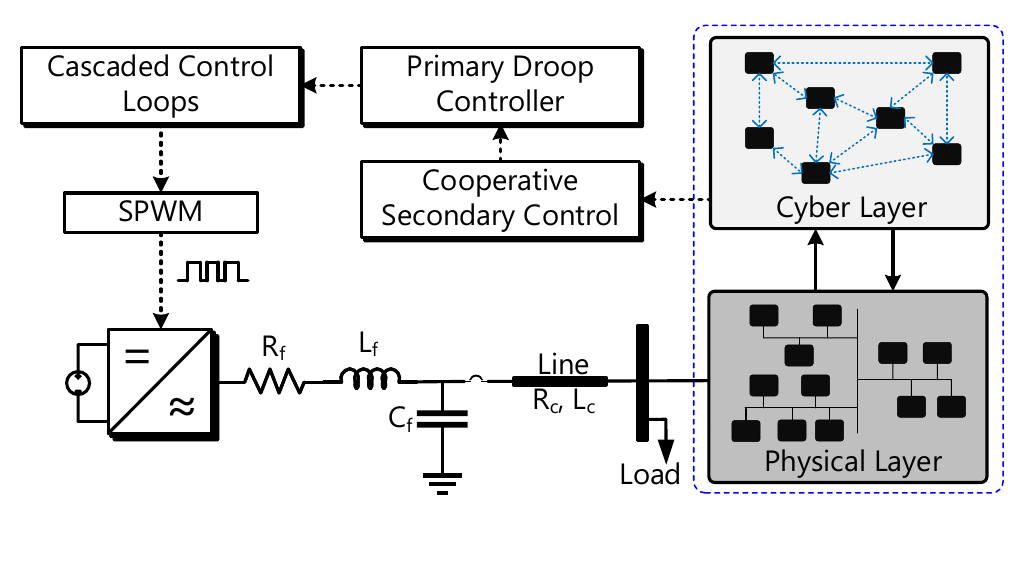}
    \caption{ \label{CONTROL} Hierarchical control in distributed microgrids.}
\end{figure}


\subsubsection{Primary Control}
The primary layer's core objective is to achieve synchronization in voltage and frequency values for the regulation of active and reactive power sharing across Distributed Generators (DGs) in the microgrid. Droop-based primary power control action can be formulated as follows:
\begin{equation}
    \omega^{\ast}_l = \omega_{nom}-D_{P_l}P_l
\end{equation}
\begin{equation}
    v^{\ast}_{l} = v_{nom}-D_{Q_l}Q_l
\end{equation}
where, $\omega^{\ast}_l$ is the frequency at $l^{th}$ DG, $v^{\ast}_l$ is voltage, and $\omega_{nom}$ and $v_{nom}$ are frequency and voltage set-points. $D_{P_l}$ and $D_{Q_l}$ signify the droop gains corresponding to active and reactive power controllers. The droop gains adhere to the following conditions in a $N$-DG microgrid:
\begin{equation}
    D_{P_1}\cdot P_1 = D_{P_2}\cdot P_2 = ... = D_{P_N}\cdot P_N = \Delta \omega_{max}
\end{equation}
\begin{equation}
    D_{Q_1}\cdot Q_1 = D_{Q_2}\cdot Q_2 = ... = D_{Q_N}\cdot Q_N = \Delta v_{max}
\end{equation}
where $\Delta \omega_{max}$ and $\Delta v_{max}$ are the largest allowable values of frequency and voltage deviation respectively. As the primary controller has a droop-based operational framework, it results in voltage and frequency values dropping as real and reactive power values increase. To restore these values to normalcy, the secondary layer adopts a cooperative synchronization-based mechanism. This is described below.

\subsubsection{Secondary Control}
The core objective of the secondary control layer is to remove the drop in parameter values created as a consequence of the primary controller's actions. To achieve this, it utilizes a set of localized distributed controllers each of which computes two feedback signals $\delta\omega$ and $\delta v$ (for its corresponding DG) via cooperative synchronization. In this mechanism, one of the DGs is assigned the role of leader with access to reference setpoints equal to nominal values of frequency and voltage. As shown in Fig. \ref{CONTROL}, each controller is capable of communicating with its neighbors via a well-connected cyber network. The overall goals for the secondary layer can be formulated as:
\begin{equation}
    \lim_{t\to \infty} ||\omega_l - \omega_n|| = 0 \;\forall\;l\label{oj1}
\end{equation}
\begin{equation}
    \lim_{t\to \infty} ||D_{P_l}P_l - D_{P_m}P_m|| = 0 \;\forall\;l,m\label{oj2}
\end{equation}
\begin{equation}
    \lim_{t\to \infty} ||D_{Q_l}Q_l - D_{Q_m}Q_m|| = 0 \;\forall\;l,m\label{oj3}
\end{equation}
To achieve these goals, the $l^{th}$ secondary controller directly feeds $\delta\omega_l$ and $\delta v_l$ to the primary power controller dynamics.
Hence, the nominal control framework in distributed microgrids achieves set-point tracking and global synchrony via the following power control dynamics at the local DG level:
\begin{eqnarray}
\omega^{\ast}_l = \omega_{nom}-D_{P_l}P_l + \delta\omega_l\label{sys1}\\
v^{\ast}_{l} = v_{nom}-D_{Q_l}Q_l + \delta v_l\label{sys2}
\end{eqnarray}
$\delta\omega$ and $\delta v$ are determined as per the following single integrator dynamics:
\[
\delta\dot{\omega_l} = K_1\Big(\sum_{m\in{N(l)}}{a_{lm}}(\omega_{m}-\omega_l)+g_l(\omega_n-\omega_l)+
\]
\begin{eqnarray}
\sum_{m\in{N(l)}}{a_{lm}}(D_{P_m}P_{l}-D_{P_l}P_{l})\Big)\label{dynamics1}\\
\delta\dot{v_l} = K_2\Big(\sum_{m\in{N(l)}}{a_{lm}}(D_{Q_m}Q_{m}-D_{Q_l}Q_{l})\Big)\label{dynamics2}
\end{eqnarray}
where, $a_{lm}$ is an element of the adjacency matrix representing the communication spanning tree $s_T[i] \in S_T$. $S_T$ is the set of spanning trees, each consisting of a unique set of network devices (e.g., transmitters, receivers, repeaters, etc). A noteworthy point is that each spanning tree in $S_T$ can achieve nominal secondary control objectives in the microgrid environment without affecting nominal stability. 
Further, spanning trees for a microgrid (or for any graphical network for that matter) are non-unique in nature \cite{gabow1978finding}. Each possible spanning tree in this context consists of a different set of communication devices (e.g., transmitters, repeaters, etc.). This means that every pair of non-unique spanning trees will have one or more non-overlapping communication elements.


\subsubsection{DIAs in the Communication Layer}

DIAs in the microgrid environment are typically executed via the cyber layer. The paper considers two DIAs, each initiated from a unique vulnerable cyber device. The first DIA involves manipulations at the transmitter level. Access to one or more transmitters in the microgrid network means that the attacker can directly falsify information at the primary source and simultaneously mislead all the controllers that make decisions based on signals from the untrustworthy transmitter(s). The second DIA is achieved via repeater-level manipulations.
In this case, the attacker can only falsify information to two different nodes simultaneously. The first one is the receiver of the recipient DG and the second one is the receiver of the transmitting DG.
This is because a single repeater often handles bidirectional communication for any given pair of communicating DGs in the microgrid network.

Each of the considered DIAs can affect the stability of the microgrid as it has a direct impact on the decisions of the secondary control layer and consequentially a cascading impact on the primary control layer.
Hence, it is important that there be a dedicated framework to identify the presence of such DIAs and neutralize/alleviate their impact on microgrid system dynamics and nominal control operations.

\section{ANN-Based Communication Graph Switching}

\begin{figure}[tp]
\centering
    \centering
    \includegraphics[width=\linewidth]{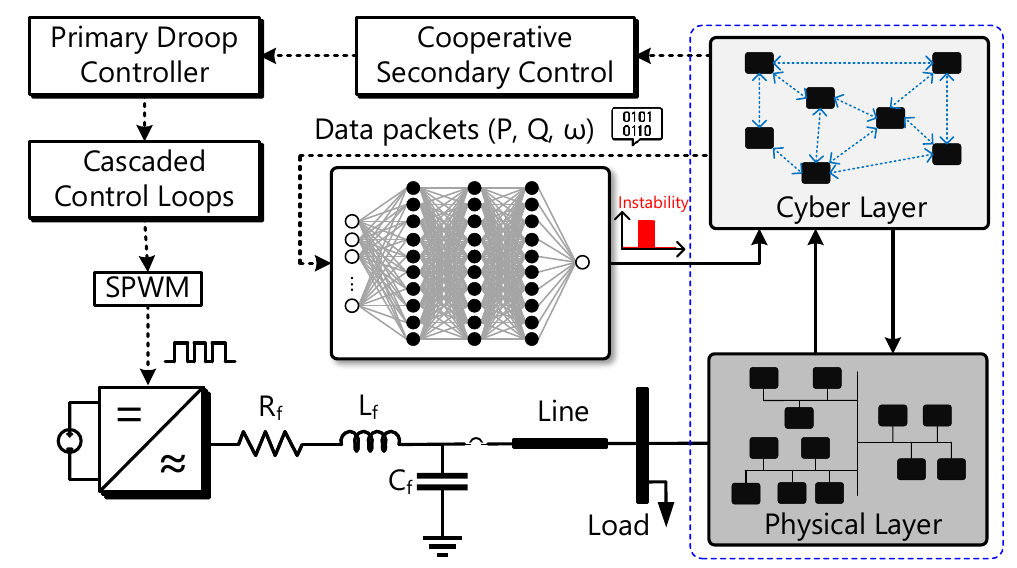}
    \caption{ \label{framework} Schematic diagram of the proposed cyberattack detection and mitigation framework.}
\end{figure}

\begin{figure*}[tp]
\centering
    \centering
    \includegraphics[width=0.963\linewidth,clip,trim={28 7 28 55}]{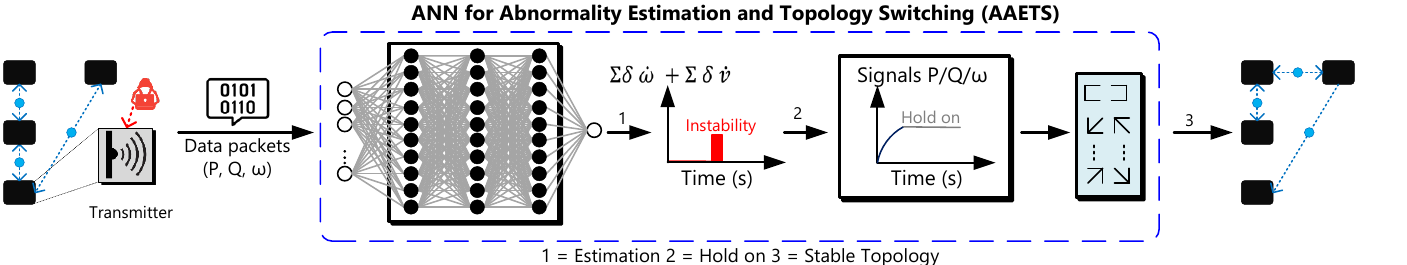}
    \caption{ \label{steps} Working mechanism of the proposed ANN-based graph topology switching framework.}
\end{figure*}

The microgrid system consists of inter-dependent physical and cyber layers. The physical layer consists of generators, load units, tie-lines, sensors, etc. The cyber layer consists of network devices that help in the exchange of information like transmitters, receivers, repeaters, etc. A crucial requirement for stability in distributed microgrids is the formation of a spanning tree at the communication network level. In the event of cyber-attacks, following a spanning tree communication network topology that is dependent on compromised network devices can lead to attack propagation and instability. We consider two attack vectors:
\begin{enumerate}
    \item FDI attacks which are executed via one or more compromised transmitters. In the network graph topology, transmitters are an integral part of the nodes. Hence, a compromised transmitter will mean that all outgoing information from the node(s) is untrustworthy. Mitigation of this vector will require that no outgoing signal from a compromised node is utilized for the achievement of control objectives.
    \item MITM attacks via compromised signal repeaters. In the microgrid communication network graph, repeaters are part of the communication link (edge) and can be exploited to achieve bidirectional manipulation of information. Mitigation of MITM attacks will require that the corresponding link is not actively involved in feeding measurement signals to the secondary controllers.
\end{enumerate}
Using compromised transmitters and/or repeaters in a communication network topology makes it untrustworthy and incapable of achieving secondary control objectives. To determine the trustworthiness of the communication graph topology, we utilize a physics-guided ANN framework (visual depiction in Fig. \ref{framework}) that is trained in a supervised manner to estimate the possibility of abnormal secondary control behavior that may be generated if it is continually used in the system.
As shown in Fig. \ref{steps}, in case the current topology is found to be untrustworthy and cyberattack-infected, the mitigation framework iterates through possible spanning tree communication graph topologies and identifies the one that preserves normalcy in the system. Prior analysis in \cite{rath2024improvise} has shown that there will always be at least one trustworthy spanning tree topology even if (N-1) DGs in an N-DG microgrid are cyber-attack-infected. This means that unless 100\% of the microgrid network is cyberattack-infected, at least one spanning tree can always achieve resilience even under the influence of active cyber manipulations.
A detailed depiction of the ANN-assisted attack detection and topology switching framework is provided in the following text.

\begin{figure}[tp]
\centering
    \centering
    \includegraphics[width=0.47\linewidth,clip,trim={17 17 17 17}]{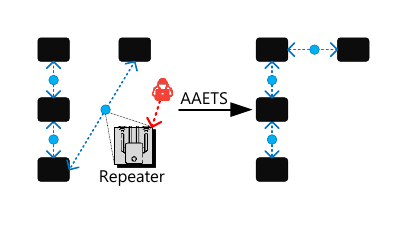}\hspace{0.41cm}
        \includegraphics[width=0.47\linewidth,clip,trim={17 17 17 17}]{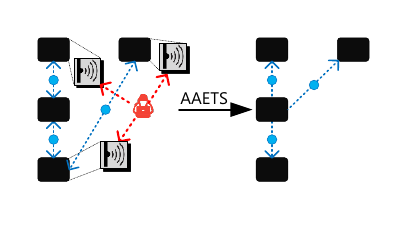}\\(a)\hspace{4.5cm}(b)
    \caption{ \label{steps1} Demonstration of (a) resilience against repeater-level MITM attacks and (b) $(N-1)$ resilience against transmitter-level FDI attacks.}
\end{figure}



\subsection{Physics-Guided Deep Learning for Cyber-Attack Indication}

		


$\delta\dot{\omega_l}$ and $\delta\dot{v_l}$ are essentially error computations between inter-DG sensor measurements. We use these terms to create a physics-guided principle for our ANN-based regression model that can estimate abnormal secondary control behavior indicating cyberattacks.
Consider a fused sum of all secondary control signals ($T_{pr})$ in the microgrid model. Mathematically, we express this as:
\begin{equation}
    T_{pr} = \Big\{\sum_{l=1}^N \delta\dot{\omega_l} + \sum_{l=1}^N \delta\dot{v_l}\Big\}
\end{equation}
From equations \ref{dynamics1} and \ref{dynamics2}, we can write $T_{pr}$ as:
\[ T_{pr} = \sum_{l=1}^N\Big(K_1\Big(\sum_{m\in{N(l)}}{a_{lm}}(\omega_{m}-\omega_l)+g_l(\omega_n-\omega_l)+\]
\[
\sum_{m\in{N(l)}}{a_{lm}}(D_{P_m}P_{l}-D_{P_l}P_{l})\Big) +
\]
\begin{equation}
K_2\Big(\sum_{m\in{N(l)}}{a_{lm}}(D_{Q_m}Q_{m}-D_{Q_l}Q_{l})\Big)\Big)
\end{equation}
In the steady state, the secondary control layer strives to achieve the objectives in equations \ref{oj1}, \ref{oj2}, and \ref{oj3}. This means that the following conditions would be satisfied:
\begin{equation}
    \omega_n \approx \omega_1 \approx ... \approx \omega_N\label{ss1}
    \end{equation}
    \begin{equation}
    D_{P_1}\cdot P_1 \approx D_{P_2}\cdot P_2 \approx ... \approx D_{P_N}\cdot P_N\label{ss2}
    \end{equation}
    \begin{equation}
D_{Q_1}\cdot Q_1 \approx D_{Q_2}\cdot Q_2 \approx ... \approx D_{Q_N}\cdot Q_N\label{ss3}
\end{equation}
Using equations \ref{ss1}, \ref{ss2}, and \ref{ss3}, we can estimate a steady state value for $T_{pr}$ as:
\begin{equation}
    T_{pr} \approx 0
\end{equation}
To replace the approximate equality in the above equation with an exact equality, we introduce an infinitesimal term $\sigma$.
\begin{equation}
    T_{pr} = \sigma \label{law}
\end{equation}
However, in the presence of FDI and MITM vectors, equation \ref{law} becomes invalid.
This is because the communicated signals $\{\omega, P, Q\}$ are modified by the attacker to incorporate a bad exogenous data signal $X_A$.
Hence, in the presence of an attack vector,
\[ T_{pr} = \sum_{l=1}^N\Big(\sum_{m\in{N(l)}}{a_{lm}}(\omega_{m}-\omega_l)+g_l(\omega_n-\omega_l)+\]
\[
\sum_{m\in{N(l)}}{a_{lm}}(D_{P_m}P_{l}-D_{P_l}P_{l})\Big) +
\]
\begin{equation}
K_2\Big(\sum_{m\in{N(l)}}{a_{lm}}(D_{Q_m}Q_{m}-D_{Q_l}Q_{l})\Big)\Big) + X_A
\end{equation}
On further simplification, we can say that during any DIA (irrespective of whether it is FDI or MITM),
\begin{equation}
    T_{pr} = \sigma + X_A
\end{equation}
Considering the inherent nature of microgrid physics, we use ANN as an estimator for $T_{pr}$. This helps it to serve as an indicator of cyberattacks in the microgrid network. Thus exploiting this physics-guided property, we use a $L$-layered deep ANN that is trained in a supervised manner to estimate a single output feature $T_{pr}$ based on all the $P, Q, \omega$ to be received by each $DG$ as per the current spanning tree $s_T[i]$. Note that these features are collected from each DG in the network.
If the value of $T_{pr}$ is found to be higher than $\sigma$, a trigger is raised indicating the presence of a cyberattack in the microgrid environment.



\begin{table}
	\renewcommand{\arraystretch}{1.3}
	\caption{{4-DG Microgrid and ANN Hyperparameters}}
	\label{sys}
	\centering
\begin{threeparttable}
	\begin{tabular}{|c|c|c|c|}
		\hline
		Parameter & Value & Parameter & Value \\
		\hline
		\hline
		\(V_{dc}\) & 1000 V  & Line & 0.5 mH + 0.07 $\Omega$ \\  
		\hline
		\(R_f\) & 0.1 $\Omega$  & $L_f$ & 4 mH \\  
		\hline
		\(R_c\) & 0.1 $\Omega$  & $C_f$ & 200 $\mu F$ \\  
		\hline
		\(\alpha\) & 0.001  & $L$ & 5 \\  
		\hline
		\(N\) & 4 & \(D_P\) & \(1\times{10^{-4}}\)  \\ 
		\hline
		\(\beta\) & 10 & \(D_Q\) & \(1\times{10^{-4}}\)  \\ 
		\hline
		\(N_{ET}\) & 5000 & \(w_{nom}\) & 50 Hz \\ 
		\hline
		 \(P_{Ep}\) &  50 &  $N_{DT}$ & 5,000,000 \\ 		
		\hline 
            $t_a$ & 5 s & \(t_{total}\) & 10 s \\
            \hline 
	\end{tabular}
\end{threeparttable}
\end{table}

\subsection{Optimal Spanning Tree Switching for Attack Mitigation}
As shown in Fig. \ref{steps}, on the receipt of the trigger, a $\texttt{Hold}$ is initiated that keeps the state of the system immune from the estimated controller abnormality due to cyberattacks. The \texttt{Hold} is retained until the ANN is made to estimate $T_{pr}$ values for all the spanning trees in $S_T$. Finally, the first spanning tree graph topology with $T_{pr} = \sigma$ is chosen as the active communication topology and nominal operations are resumed again.
This preserves system stability in the presence of the cyberattack and removes the impact of the attack from the microgrid dynamics before it has a chance to affect the system.
As depicted in Fig. \ref{steps1} (a), this framework can also identify and isolate repeater-level DIAs.
A noteworthy point is that the framework can achieve resiliency even if $N-1$ transmitters in the microgrid network are cyberattack-infected. This is also depicted in Fig. \ref{steps1} (b).

\begin{table}
    \fontsize{9}{13}\selectfont
    \renewcommand{\arraystretch}{1.3}
    \caption{Performance of the Proposed Abnormality Estimation Model}
    \label{performance}
    \centering
    \begin{threeparttable}
    \begin{tabular}{|c|c|c|c|c|}
        \hline
        $\text{SNR}_\text{dB}$ & Performance & Training & Validation & Testing \\
        \hline
        \hline
        
        {$\infty$} & MAE & 0.01136 & 0.01137 & 0.01138 \\  
        \cline{2-5}
        & MSE & 0.0002 & 0.0002 & 0.0002  \\ 
        \cline{2-5}
        & RMSE & 0.01416 & 0.01419 & 0.01418  \\ 
        \hline

        
        {75 dB} & MAE & 0.01175  & 0.01176 & 0.01176 \\  
        \cline{2-5}
        & MSE & 0.00022 & 0.00023 & 0.00023  \\ 
        \cline{2-5}
        & RMSE & 0.01499 & 0.01504 & 0.01502  \\ 
        \hline
        {70 dB} & MAE & 0.00964  & 0.00965 & 0.00965 \\  
        \cline{2-5}
        & MSE & 0.00017 & 0.00017 & 0.00017  \\ 
        \cline{2-5}
        & RMSE & 0.01316 & 0.01321 & 0.01320  \\ 
        \hline
        {65 dB} & MAE & 0.01223  & 0.01224 & 0.01224 \\  
        \cline{2-5}
        & MSE & 0.00024 & 0.00024 & 0.00024  \\ 
        \cline{2-5}
        & RMSE & 0.01546 & 0.01550 & 0.01549  \\
        \hline
        {60 dB} & MAE & 0.01257  & 0.01258 & 0.01258 \\  
        \cline{2-5}
        & MSE & 0.00025 & 0.00025 & 0.00025  \\ 
        \cline{2-5}
        & RMSE & 0.01590 & 0.01592 & 0.01592  \\ 
        \hline
        {55 dB} & MAE & 0.01314  & 0.01315 & 0.01315 \\  
        \cline{2-5}
        & MSE & 0.0003 & 0.0003 & 0.0003  \\ 
        \cline{2-5}
        & RMSE & 0.01735 & 0.01737 & 0.01737  \\
        \hline
        {50 dB} & MAE & 0.12979  & 0.1300 & 0.12994 \\  
        \cline{2-5}
        & MSE & 0.03975 & 0.03984 & 0.03985  \\ 
        \cline{2-5}
        & RMSE & 0.19938 & 0.19960 & 0.19963  \\ 
        \hline
        {45 dB} & MAE & 0.01936  & 0.01933 & 0.01935 \\  
        \cline{2-5}
        & MSE & 0.00065 & 0.00065 & 0.00065  \\ 
        \cline{2-5}
        & RMSE & 0.02547 & 0.02544 & 0.02545  \\ 
        \hline
        {40 dB} & MAE & 0.06748  & 0.06745 & 0.06749 \\  
        \cline{2-5}
        & MSE & 0.00496 & 0.00496 & 0.00497  \\ 
        \cline{2-5}
        & RMSE & 0.07046 & 0.07044 & 0.07047  \\ 
        \hline
    \end{tabular}
    \end{threeparttable}
\end{table}

\subsection{Rationale Behind the Proposed Method}
The rationale behind the proposed abnormality estimation method is derived from \cite{sahoo2018stealth} which highlights that stealthy DIA attacks lead to the disruption of consensus among microgrid DGs diverging one or more secondary control outputs from their nominal value which is approximately 0.
The method presented in \cite{sahoo2018stealth} involves attack detection only after local secondary controllers have processed them indicating cyberattack progression from the communication layer to the secondary control plane. This can lead to a higher risk of increased time delay and/or instability as the attack has already achieved a certain degree of penetration within the control plane. However, the regression mechanism in our paper attempts to estimate possible secondary controller-level abnormality that can be indicative of DIAs even before the attack progresses from the communication/network layer to the control plane thereby attempting to lower the risk of instability and achieve mitigation with minimal time delays.
Furthermore, our method estimates a fused sum of all individual secondary control outputs within the microgrid environment meaning that any DIA irrespective of its target and mode of propagation is identified via a unified framework represented by the ANN outputting an estimation for $T_{pr}$.
Then, a hold is introduced, and the active network graph topology is switched from the current topology to another pre-defined spanning tree topology whose estimation for $T_{pr}$ conforms to equation \ref{law}. As per \cite{gabow1978finding}, there can be several such topologies for any given (microgrid) graph. Each of them would still lead to the achievement of secondary control objectives within the microgrid.







\section{Performance Validation and Results}

\begin{figure}
    \centering
    \begin{minipage}{\linewidth}
        \centering
        \includegraphics[width=\linewidth]{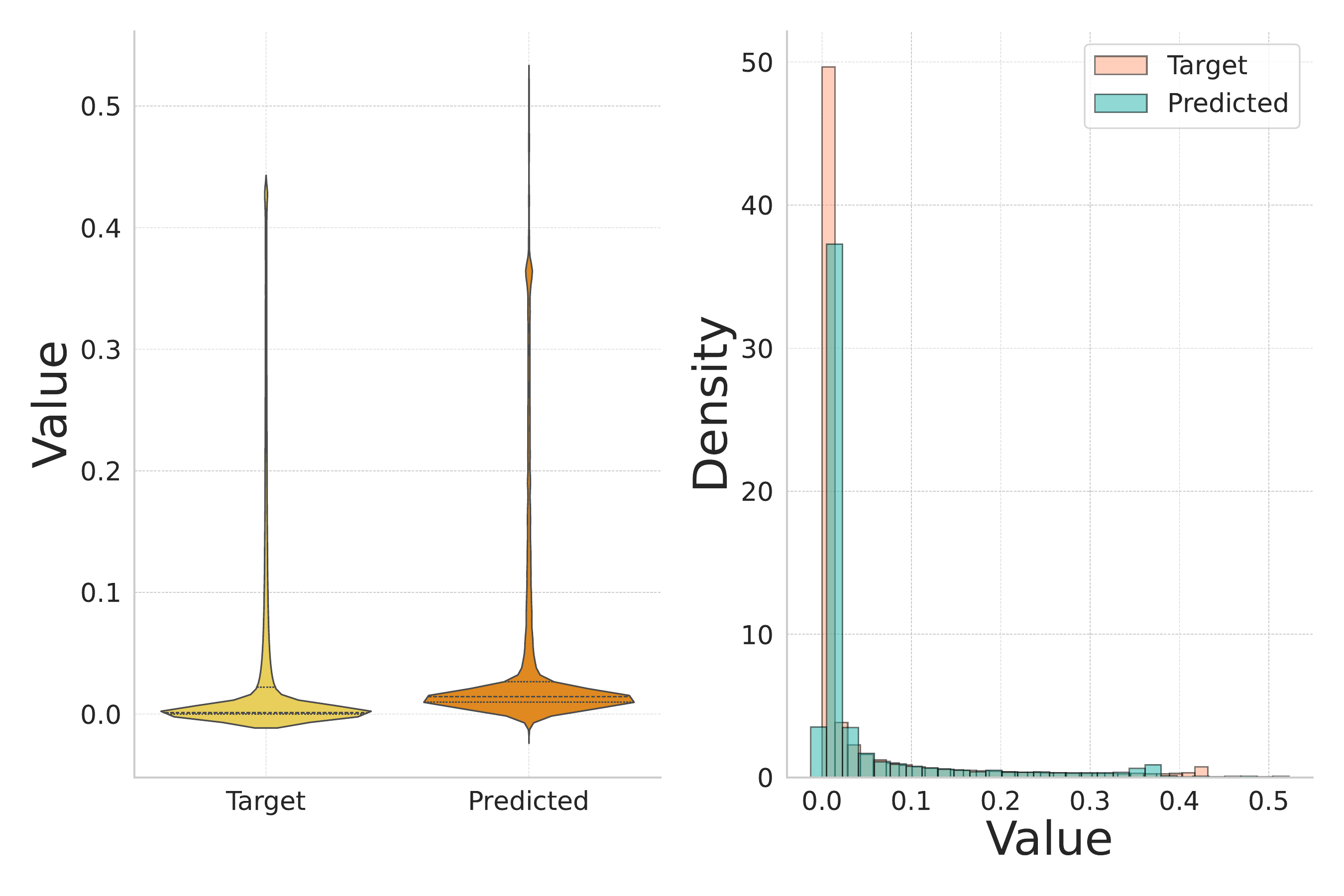}\\(a)\hspace{3.7cm}(b)
    \end{minipage}%
    \hfill
    \caption{Performance of the deep ANN model without noise: (a) violin plot showing target and predicted value distributions, and (b) density histogram comparing target and predicted value frequencies.}
    \label{fig:ANN_noSNR}
\end{figure}

\begin{figure}
    \centering
    \begin{minipage}{\linewidth}
        \centering
        \includegraphics[width=\linewidth]{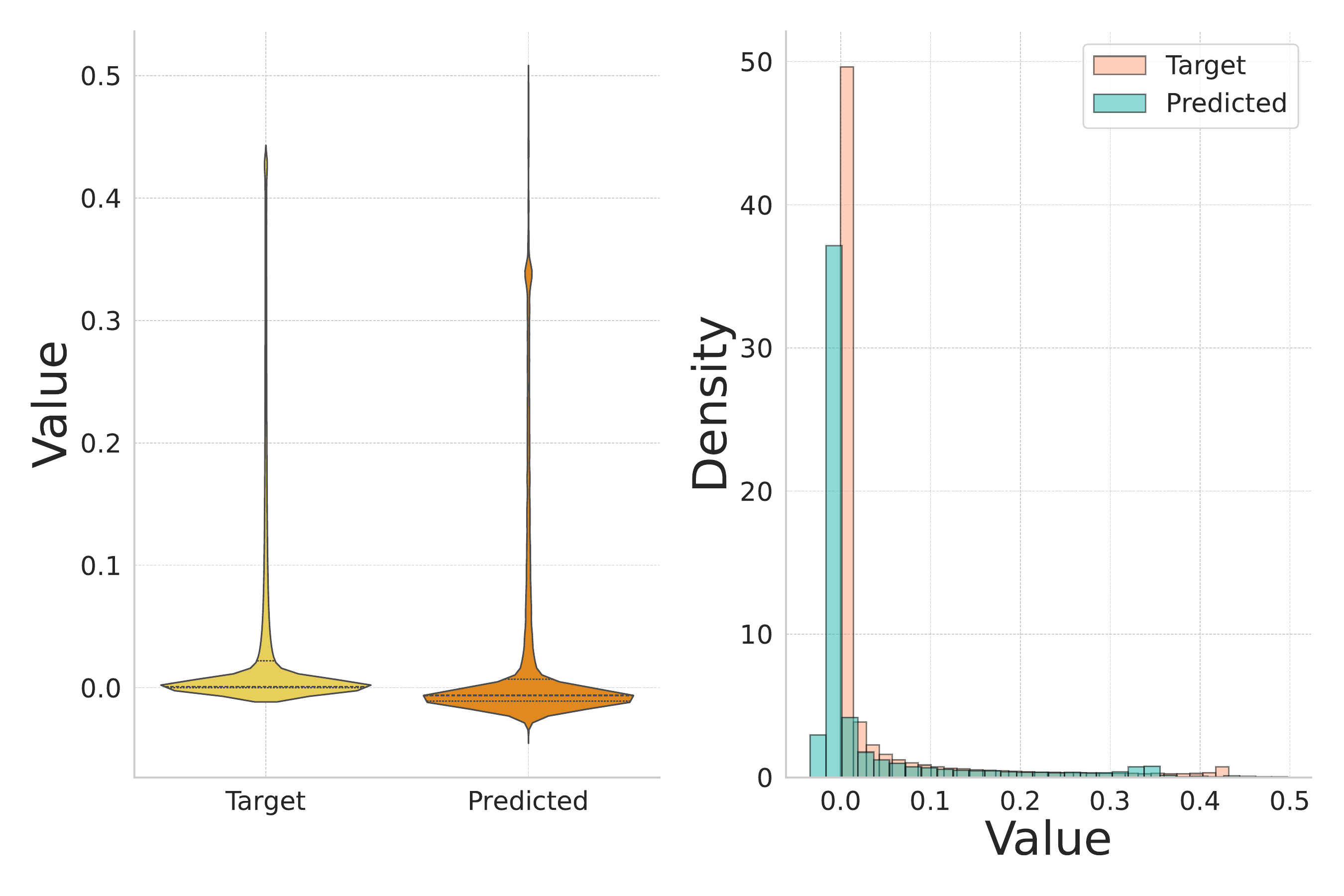}\\(a)\hspace{3.7cm}(b)
    \end{minipage}%
    \hfill
    \caption{Performance with a noisy dataset of $\text{SNR}_\text{dB}$ = 75 dB: (a) target and predicted value distributions, and (b) density histogram showing limited overlap between target and predicted values.}
    \label{fig:ANN_75}
\end{figure}


\begin{figure}
    \centering
    \begin{minipage}{\linewidth}
        \centering
        \includegraphics[width=\linewidth]{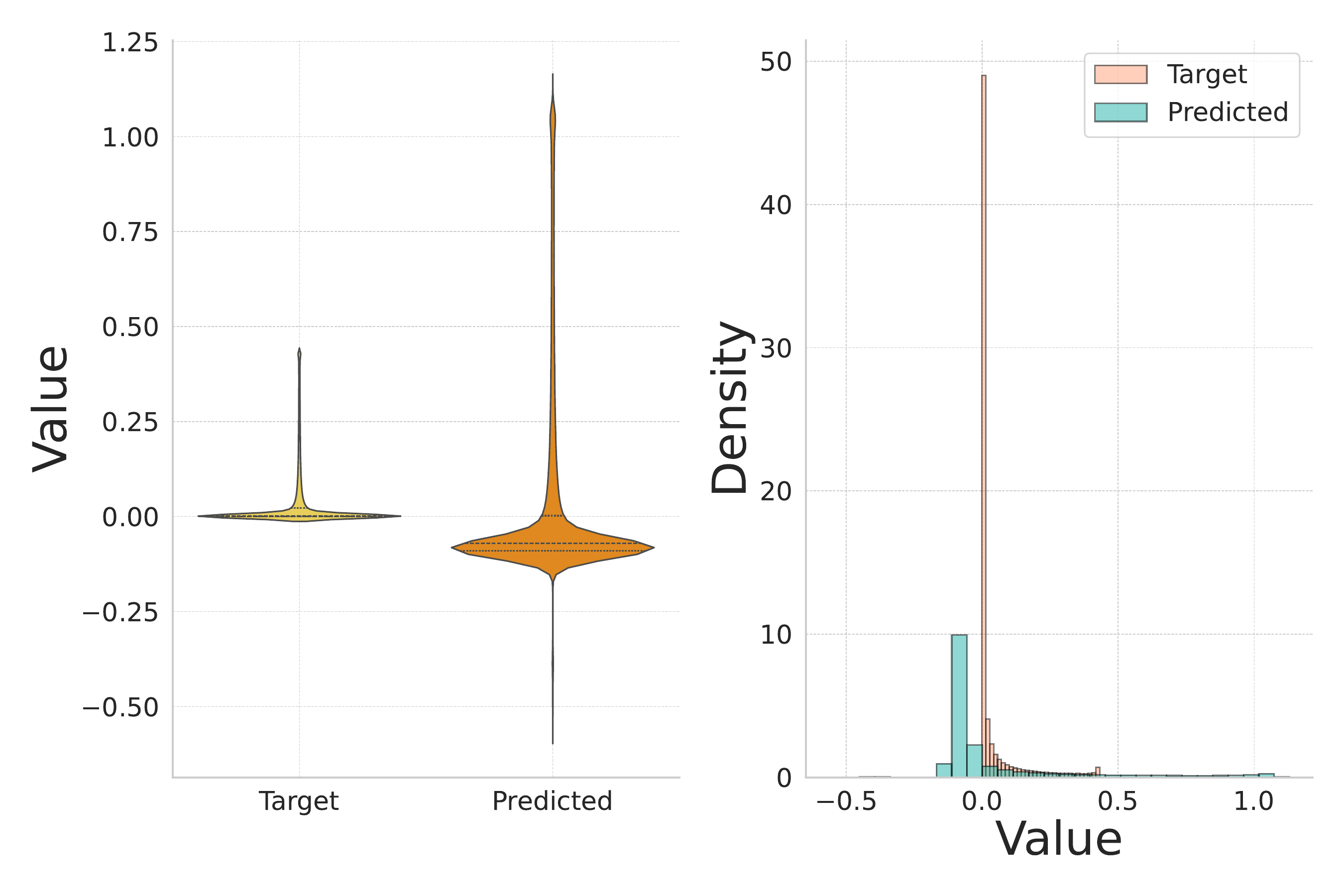}\\(a)\hspace{3.7cm}(b)
    \end{minipage}%
    \hfill
    \caption{Performance with a high-noise dataset of $\text{SNR}_\text{dB}$ = 50 dB (a) significant misalignment between target and prediction distributions and (b) minimal overlap between target and predicted value densities.}
    \label{fig:ANN_50}
\end{figure}




\begin{table}
	\renewcommand{\arraystretch}{1.3}
	\caption{{Key Parameters for 10-DG Microgrid}}
	\label{sys_scalable}
	\centering
\begin{threeparttable}
	\begin{tabular}{|c|c|c|c|}
		\hline
		Parameter & Value & Parameter & Value \\
		\hline
		\hline
		\(V_{dc}\) & 1015 V  & Line & 0.5 mH + 0.09 $\Omega$ \\  
		\hline
		\(R_f\), \(R_c\) & 0.1 $\Omega$  & $L_f$ & 4 mH \\  
		\hline
		\(N\) & 10  & $C_f$ & 200 $\mu F$ \\  
		\hline
		\(w_{nom}\) & 50 Hz & \(D_P\), \(D_Q\) & \(1\times{10^{-4}}\)  \\ 
		\hline
	\end{tabular}
\end{threeparttable}
\end{table}

\begin{table}
    \fontsize{9}{13}\selectfont
    \renewcommand{\arraystretch}{1.3}
    \caption{Scalability of the Abnormality Estimation Model}
    \label{scalable_table}
    \centering
    \begin{threeparttable}
    \begin{tabular}{|c|c|c|c|c|}
        \hline
        $\text{SNR}_\text{dB}$ & Performance & Training & Validation & Testing \\
        \hline
        \hline
        
        {$\infty$} & MAE & 0.07096 & 0.07097 & 0.07088 \\  
        \cline{2-5}
        & MSE & 0.01792 & 0.01786 & 0.01784  \\ 
        \cline{2-5}
        & RMSE & 0.13385 & 0.13366 & 0.13357  \\ 
        \hline

        {75 dB} & MAE & 0.21574  & 0.21578 & 0.21559 \\  
        \cline{2-5}
        & MSE & 0.07029 & 0.07005 & 0.07  \\ 
        \cline{2-5}
        & RMSE & 0.26513 & 0.26468 & 0.26454  \\ 
        \hline
        {70 dB} & MAE & 1.4174  & 1.4197 & 1.41918 \\  
        \cline{2-5}
        & MSE & 3.33253 & 3.3441 & 3.34378  \\ 
        \cline{2-5}
        & RMSE & 1.82552 & 1.82869 & 1.8286  \\ 
        \hline
        {65 dB} & MAE & 1.43417  & 1.43648 & 1.43591 \\  
        \cline{2-5}
        & MSE & 3.3661 & 3.37739 & 3.37706  \\ 
        \cline{2-5}
        & RMSE & 1.83469 & 1.83777 & 1.83768  \\
        \hline
    \end{tabular}
    \end{threeparttable}
\end{table}

\begin{figure*}
    \centering
    \includegraphics[width=0.3\linewidth]{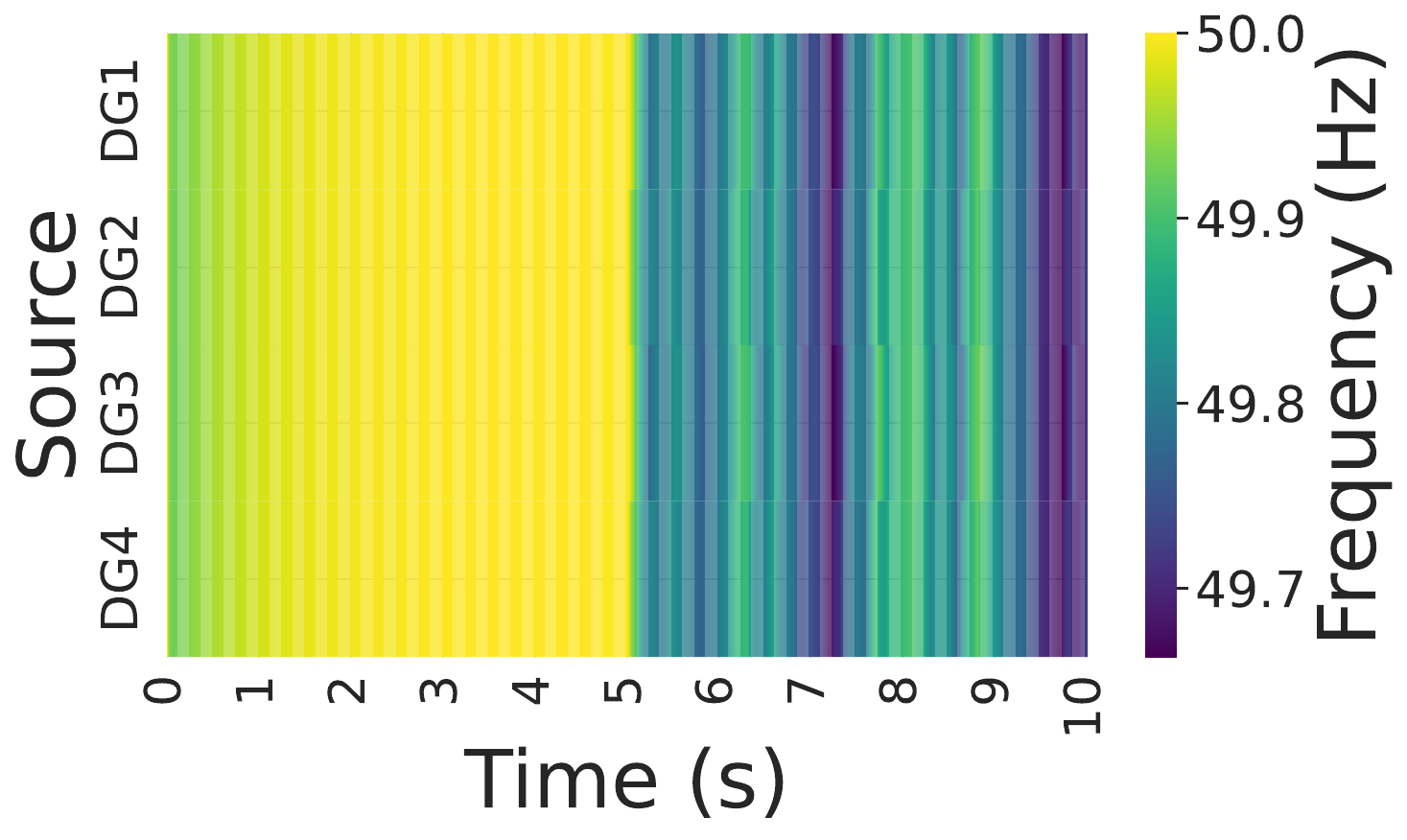}\hspace{0.51cm}
    \includegraphics[width=0.3\linewidth]{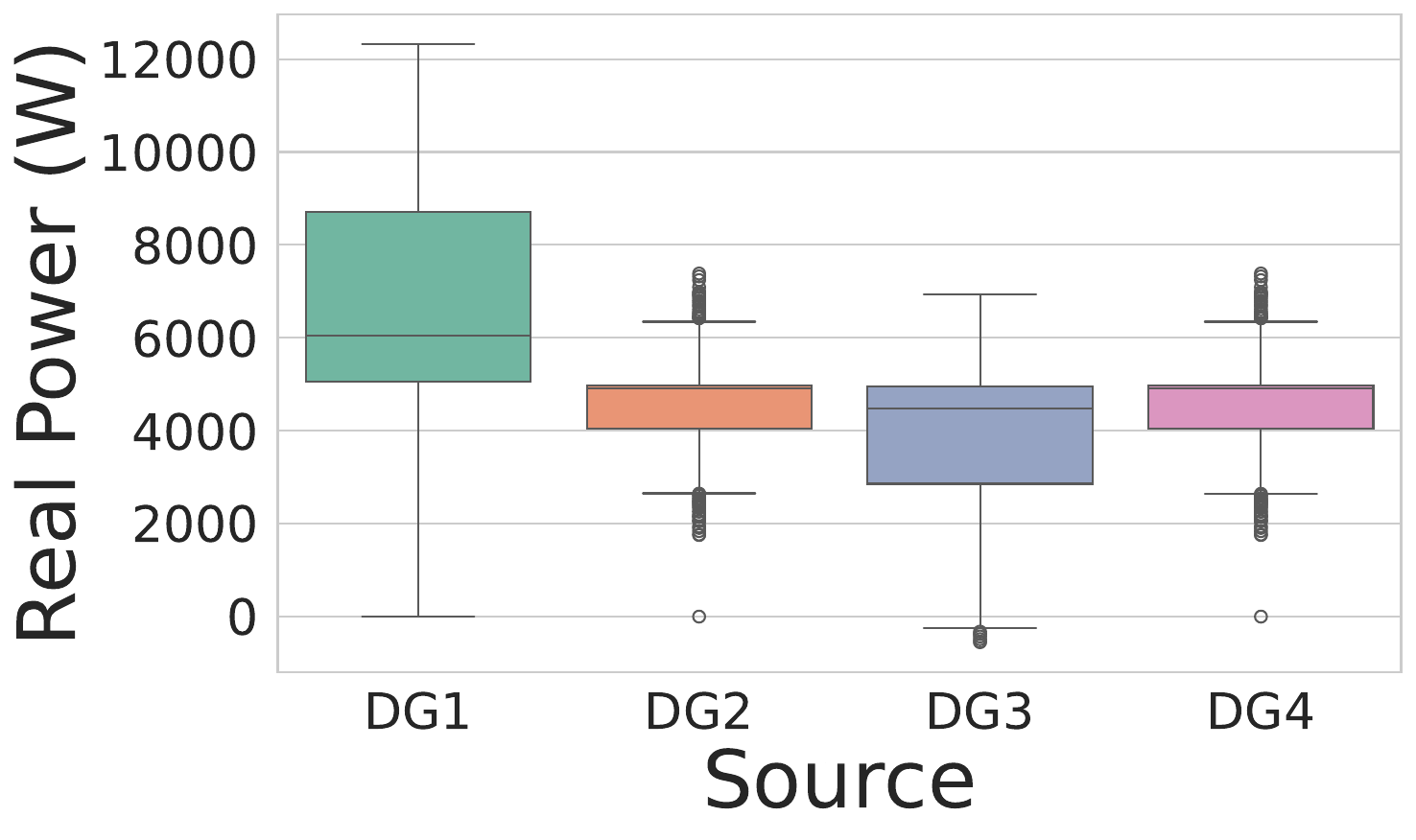}\hspace{0.51cm}
    \includegraphics[width=0.3\linewidth]{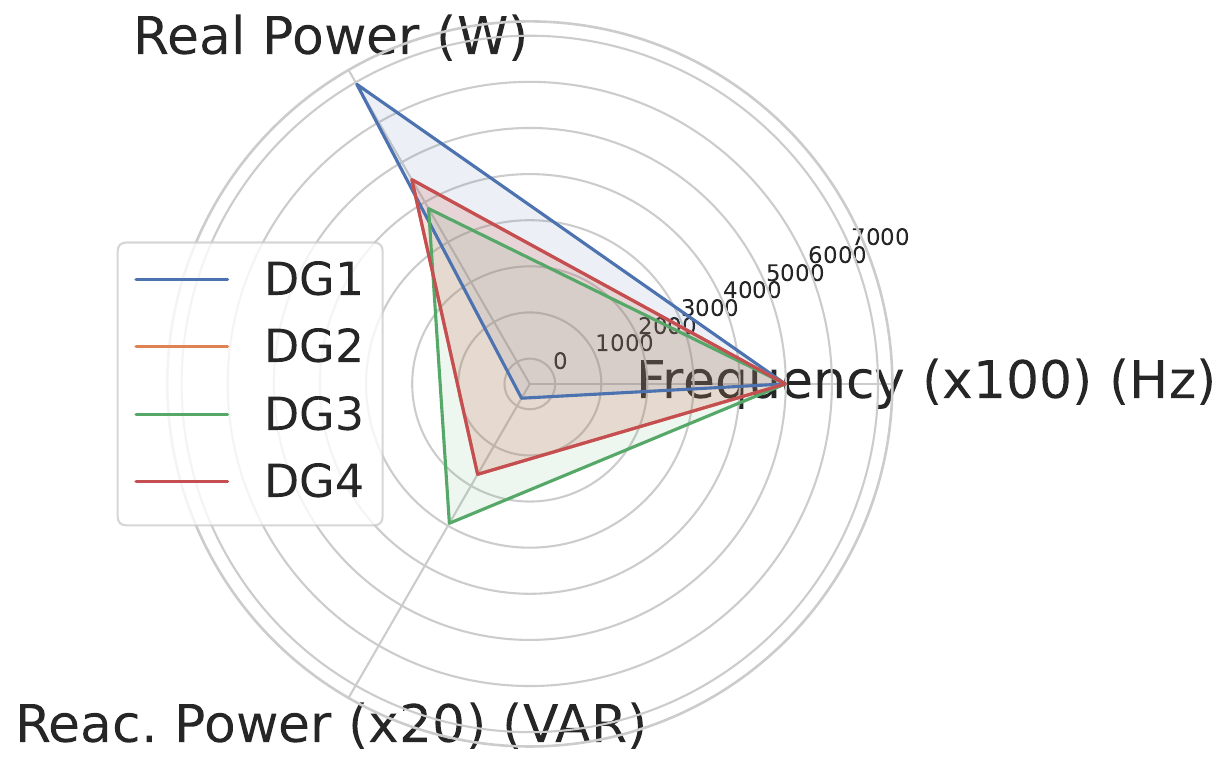}\\(a)\hspace{6cm}(b)\hspace{5cm}(c)
    \caption{{A generic microgrid does not achieve control objectives under stealth attacks. Shown here are (a) local frequency values across DGs, (b) load-sharing fluctuations, and (c) disruptions in load-sharing and frequency stability across DGs.}}
    \label{fig:attack}
\end{figure*}

\begin{figure*}
    \centering
    \includegraphics[width=0.3\linewidth]{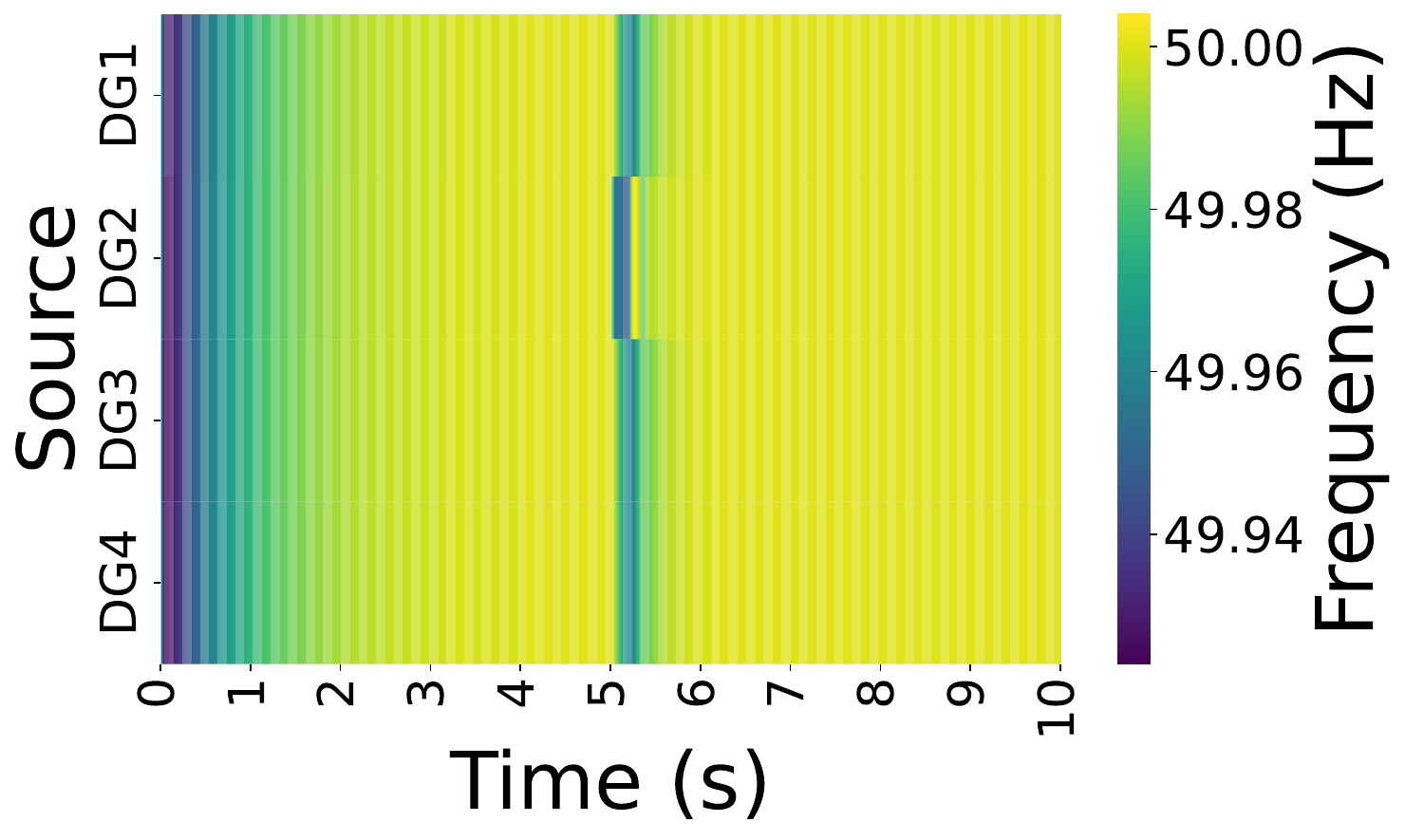}\hspace{0.51cm}
    \includegraphics[width=0.3\linewidth]{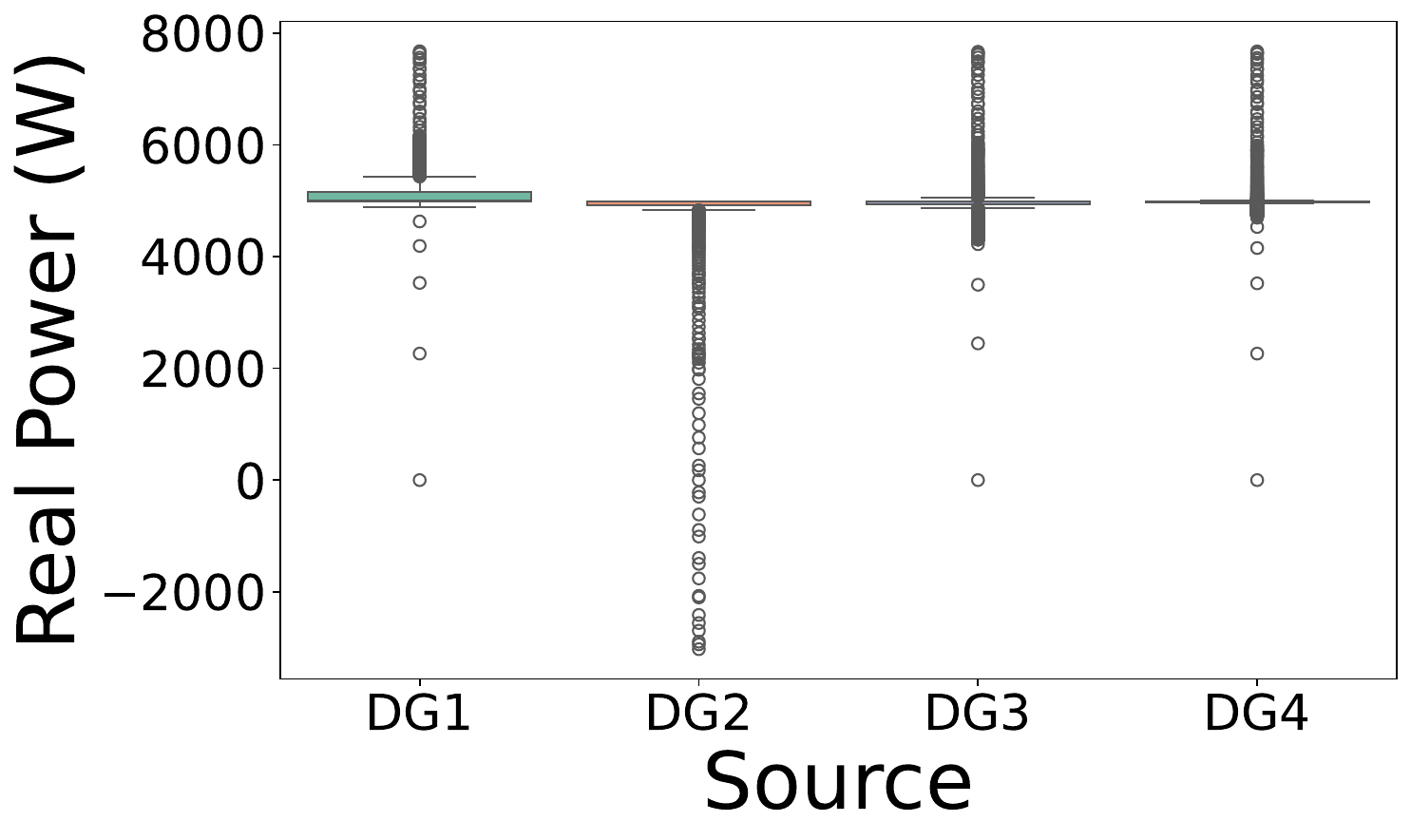}\hspace{0.51cm}
    \includegraphics[width=0.3\linewidth]{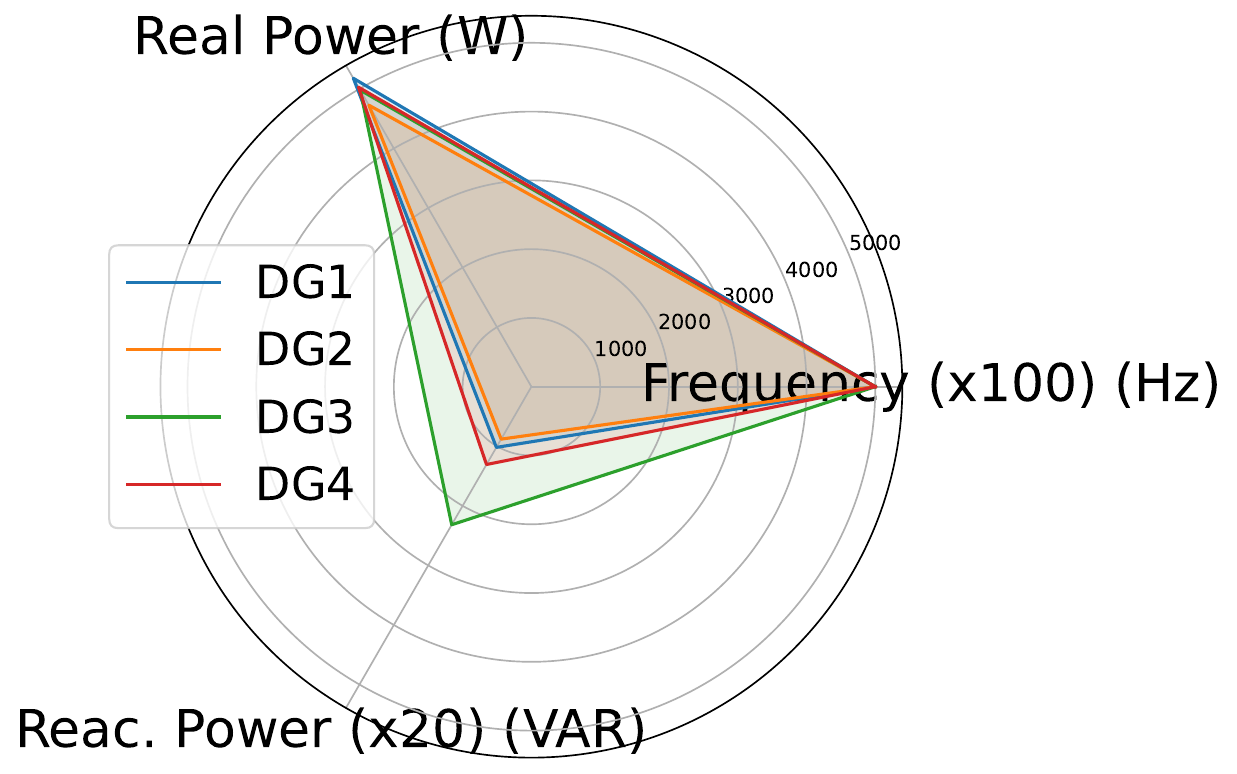}\\(a)\hspace{6cm}(b)\hspace{5cm}(c)
    \caption{{Performance of the proposed framework under FDI cyberattacks: (a)} system frequencies attain nominal values after initial disruption, (b) real power loads at DGs are consistent barring minor extremities, and (c) load sharing and frequency levels are consistent and stable.}
    \label{fig:mitigate}
\end{figure*}

\begin{figure*}
    \centering
    \includegraphics[width=0.3\linewidth]{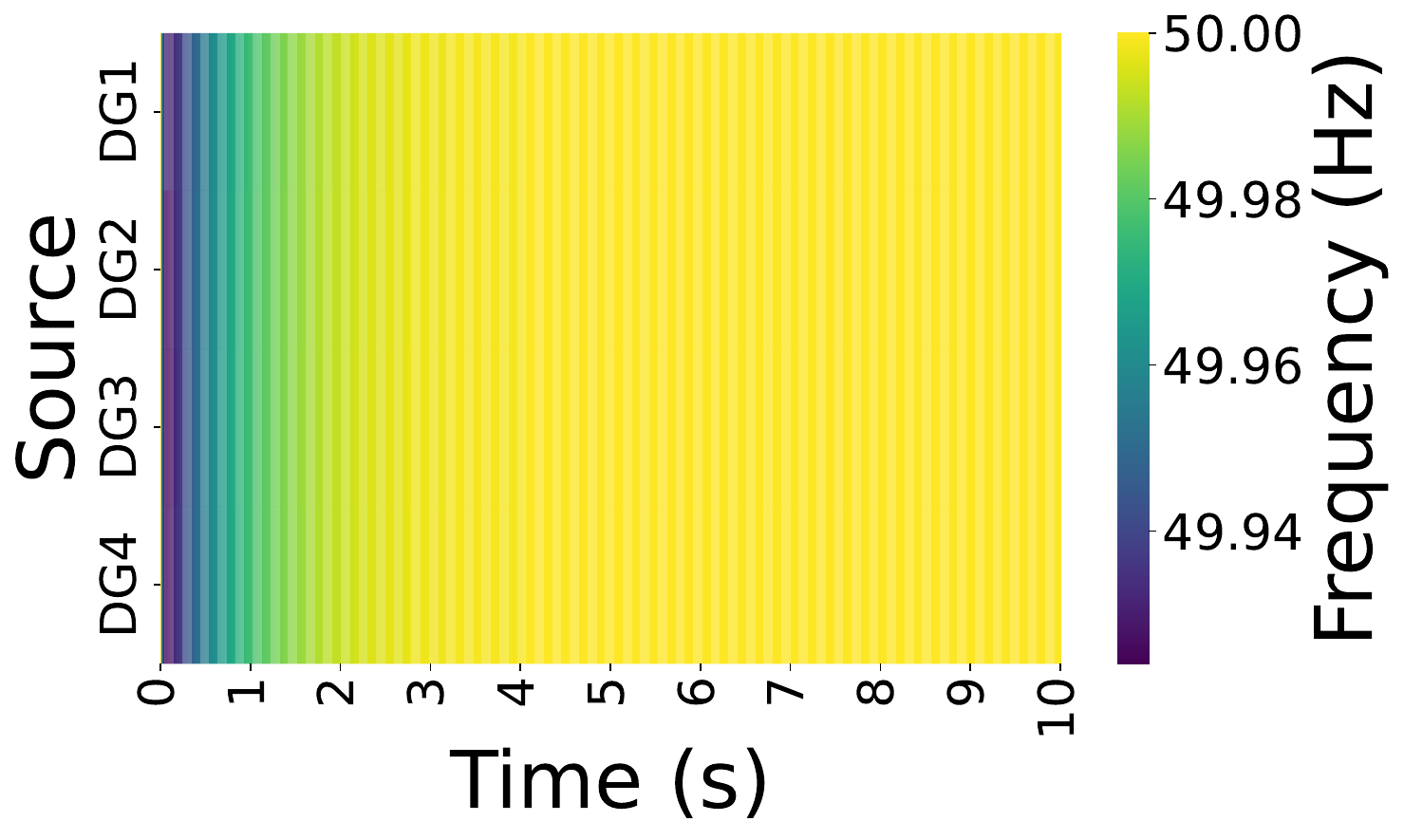}\hspace{0.51cm}
    \includegraphics[width=0.3\linewidth]{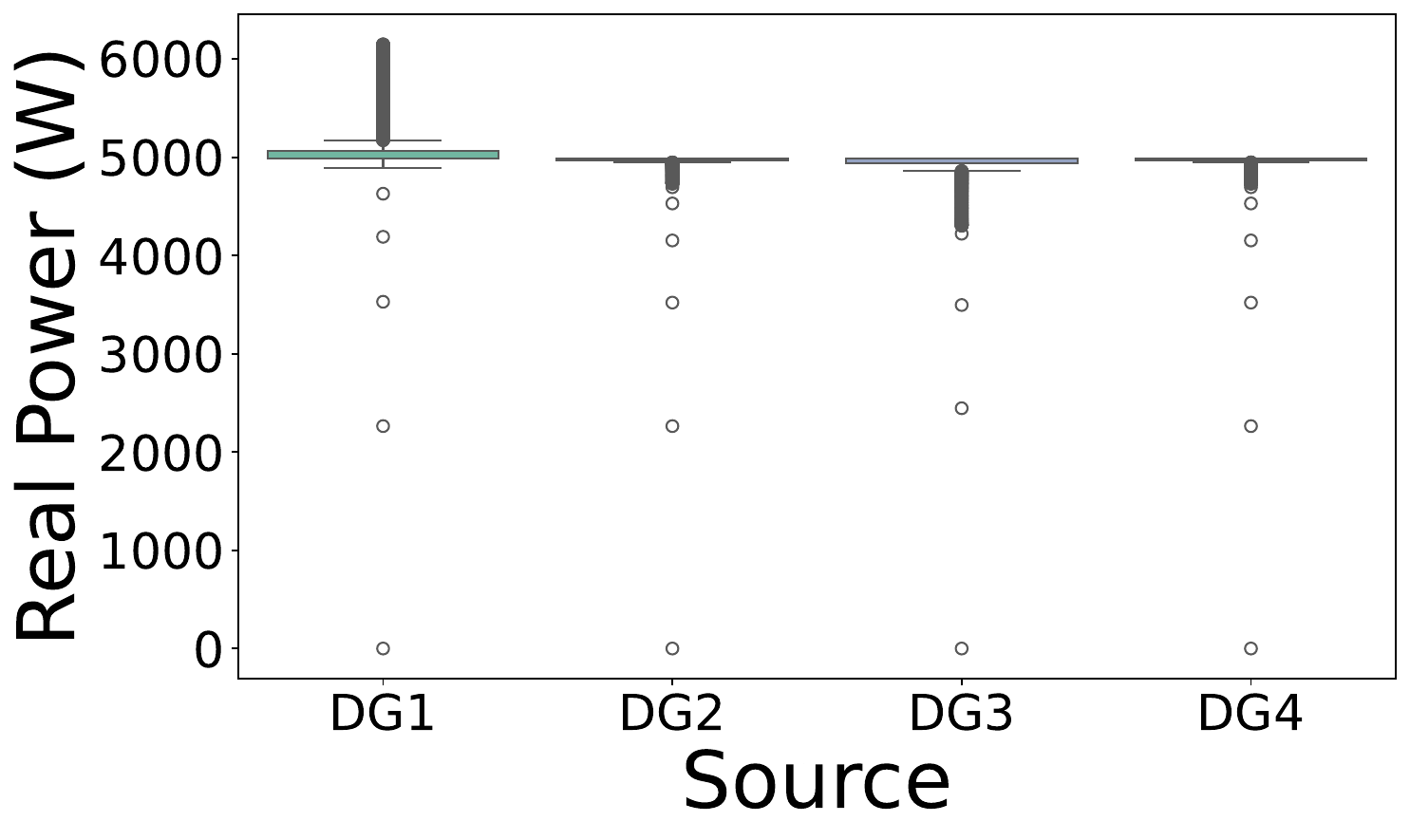}\hspace{0.51cm}
    \includegraphics[width=0.3\linewidth]{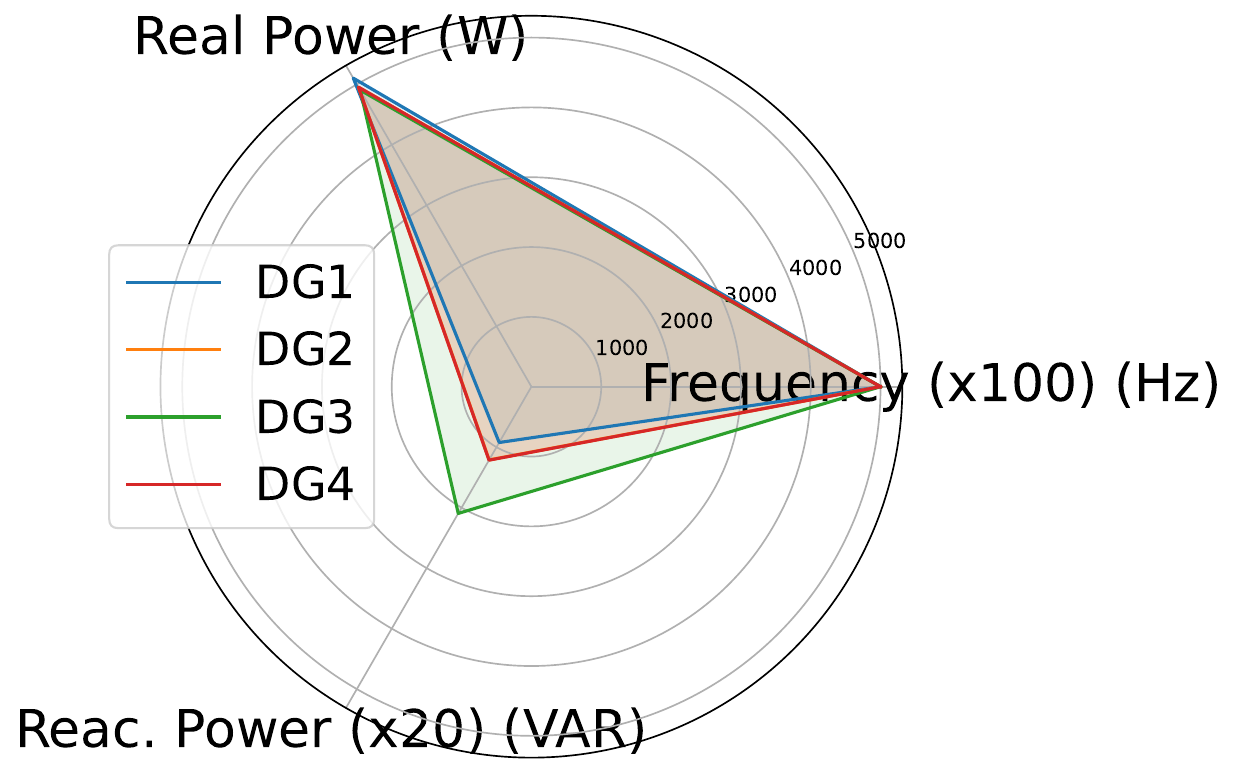}\\(a)\hspace{6cm}(b)\hspace{5cm}(c)
    \caption{{Performance under coordinated MITM attacks: (a) the impact of the attack creates no significant deviations in DG-frequency levels, (b) load levels at DGs are consistent, and (c) power-sharing follows nominal patterns.}}
    \label{fig:coordinated_mitm}
\end{figure*}

We use a MATLAB-based $N$-DG autonomous AC microgrid model for performance validation of our proposed abnormality estimation and cyberattack mitigation framework. This system follows the control architecture in Section II. Key parameters for this test system are provided in Table \ref{sys}.
The default communication graph topology for this $N$-DG model is shown in Fig. \ref{steps}. The microgrid abnormality estimation framework consists of a $L$-layered neural network (including the input, output, and three hidden layers).
Each hidden layer has $\beta$ neurons and uses the Rectified Linear Unit (ReLU) as the activation function.
The hidden layers enable the learning of complex patterns within training datasets. The loss function is MSE.
The optimizer is Adam and the learning rate is set as $\alpha$ for updating weights while training.
The maximum number of training epochs is set to $N_{ET}$.
However, we also implement and incorporate early stopping with a patience of $P_{Ep}$ epochs which means that the training process will stop if validation loss does not decrease for more than $P_{Ep}$ epochs consecutively. This is done to avoid overfitting.\\
Note that: our preliminary raw dataset consists of $N_{DT}$ data points. The data points are randomly split for training, validation, and testing purposes. We perform the splitting in two phases. In the first split, 80\% of the data is allocated for training and validation. The remaining 20\% is reserved for testing.
In the second step, we explicitly segregate the training and validation data. Here we reserve 80\% of the training/validation points for training and 20\% for the validation.
The validation set is important as it helps prevent overfitting by implementing an early stopping mechanism during training.

\subsection{Performance of estimation model}
As our preliminary raw dataset was obtained from a MATLAB-based microgrid system, it may not fully capture the noise level seen in practical, real-world datasets. To address this issue, we artificially infused the raw datasets collected from the MATLAB environment with varying noise levels. This helped us emulate practical datasets that encounter noise at the communication layer. Noise infusion was performed in a structured manner by specifying the signal-to-noise ratio in decibels ($\text{SNR}_\text{dB}$) which is defined as:
\begin{equation}
    \text{SNR}_\text{dB} = 10 \cdot \text{log}_{10} \left(\frac{P_{signal}}{P_{noise}}\right)\label{SNR}
\end{equation}
where $P_{signal}$ is the average signal power that is the squared values of the signal in consideration, $P_{noise}$ is the power of the noise to be added which is determined by rearranging equation \ref{SNR}. The magnitude of $\text{SNR}_\text{dB}$ is user-specified. $P_{noise}$ is inversely proportional to the magnitude of SNR.
The final noise to be added is sampled from a Gaussian distribution with a standard deviation equal to $\sqrt{P_{noise}}$.
Table \ref{performance} shows the Mean Absolute Error (MAE), Mean Squared Error (MSE), and Root Mean Squared Error (RMSE) values during this abnormality estimator model's training, validation, and testing, indicating its robust performance irrespective of $\text{SNR}_\text{dB}$ value. Our findings mostly follow the general trend that as $\text{SNR}_\text{dB}$ values decrease, performance deteriorates. This is also seen in the violin and density plots in Fig. 5-7. In Fig. 5, we observe significant alignment and overlap between target and estimated values during the testing phase. However, this overlap reduces when $\text{SNR}_\text{dB}$ is reduced to 75 dB (Fig. \ref{fig:ANN_75}). As $\text{SNR}_\text{dB}$ is further reduced to 50 dB (Fig. \ref{fig:ANN_50}), it observed that the overlap reduces to a higher extent indicating further performance reduction. At $\text{SNR}_\text{dB}$ = 50 dB, it is observed that some estimated abnormality values show a significant deviation from the extremities in the targets. A noteworthy point, as shown in Table \ref{performance}, is that some higher SNR$_\text{dB}$ values do not necessarily contribute to performance reduction. However, this is rare and the majority of considered cases conform to the general trend.


\subsection{Impact of cyberattacks}

\textit{\textbf{Without proposed fortification strategy:}} FDI attacks are initiated from the transmitters associated with DGs 1, 2, and 3. Fig. \ref{fig:attack} shows the impact of this attack vector in the absence of the proposed defense framework. The vector creates a deviation of frequency from its steady state and disrupts nominal power-sharing arrangements.

\textbf{\textit{Proposed ANN-based defense:}} We consider two sub-cases of attacks for performance evaluation: (i) transmitter-level FDIs and (ii) repeater-level MITMs (both initiated at $t = t_a$). During the above FDI attack strategy, in the presence of proposed defense, the physics-guided ANN analyzes the measurements flowing as per the current cyber (communication) graph and estimates that it can lead to abnormal secondary control behavior. Then a flag is raised indicating attack detection and a hold is placed on the system states. 
Then the ANN iterates through the set of pre-defined spanning trees estimating $T_{pr}$ values for them.
Finally, it identifies and enforces the topology with $T_{pr}$ conforming to equation \ref{law}.
The new topology does not rely on measurements from DGs 1, 2, and 3. This means that it will not require transmitters 1, 2, and 3 to achieve nominal functionality. 
As depicted in Fig. \ref{fig:mitigate}, local frequencies attain normalcy even in the presence of the attack vector. Power-sharing arrangements return to normal and frequency returns to 50 Hz after a minor disruption.
The second sub-case involves injecting repeater-level MITM manipulations in the links connecting DGs 2 to 3 and 3 to 4.
The framework can also achieve normalcy during this attack (Fig. \ref{fig:coordinated_mitm}).
Local frequencies and power-sharing arrangements remain unaffected even in the presence of the attack vector.

\begin{table*}
    \fontsize{9}{13}\selectfont
	\renewcommand{\arraystretch}{1.3}
	\caption{Comparative Evaluation: Proposed Approach vs. State-of-The-Art.}
	\label{comparison}
	\centering
    \begin{threeparttable}
	\begin{tabular}{|c|c|c|c|c|}
		\hline
		Basis & \cite{beikbabaei2024mitigating} & \cite{rath2024improvise} & \cite{taher2023enhancing} & \textbf{Proposed Approach} \\
		\hline
		\hline

  Deep Learning Model & LSTM & DRL & ANN & ANN \\  
		\hline

    Attack types studied & Only FDI & Rootkits & Only FDI & FDI and MITM \\  
		\hline
		
		Max. Mitigation Time & Not explicit & approx. 0.1 s & More than 1 s & Between 0.1 to 0.5 s \\  
		\hline
		Max. Resiliency against FDI & Not explicit & $(N-1)$  & Not studied  & $(N-1)$ \\ 
		\hline
  Typical Training & Slow, data expensive & Very expensive & Fast  & Fast \& less expensive \\ 
		\hline
	\end{tabular}
    \end{threeparttable}
\end{table*}




\begin{figure}
    \centering
    \includegraphics[width=0.49\linewidth]{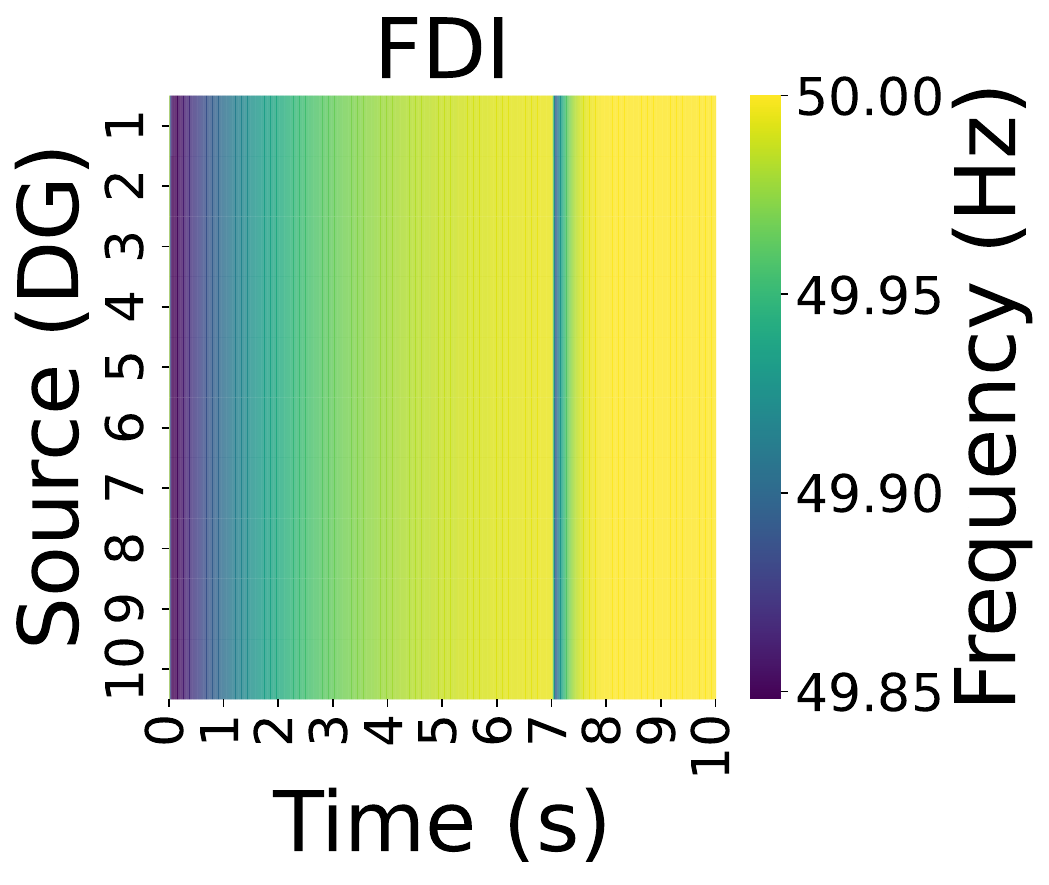}
        \includegraphics[width=0.49\linewidth]{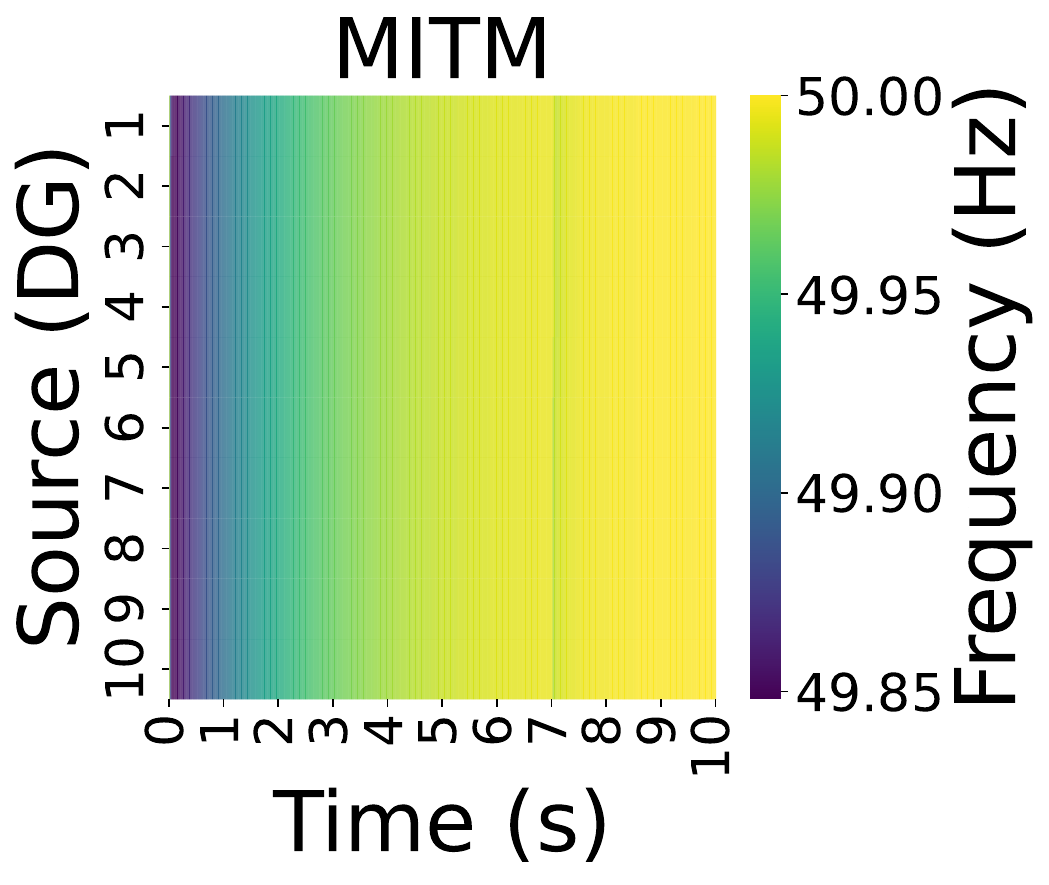}\\(a)\hspace{4.1cm}(b)
    \caption{{Performance of the framework when scaled to a 10-DG microgrid in the presence of (a) FDI attacks, and (b) MITM attacks.}}
    \label{fig:scalable}
\end{figure}

\subsection{Scalability analysis}

Microgrid sizes in the real world are not identical. Hence, the scalability of the proposed ANN model must be evaluated for higher microgrid sizes to evaluate its practical feasibility.
To perform this evaluation, we develop a 10-DG AC microgrid with system parameters as listed in Table \ref{sys_scalable}.
First, to understand how ANN performance is affected by the increased microgrid size, we retrieve data from the microgrid model and use it to train, validate, and test the $L$-layered ANN described above.
Initially, the performance evaluation is done without adding any noise. Then, synthetic noise is infused into the datasets at three distinct SNR{$_\texttt{dB}$} levels: 75 dB, 70 dB, and 65 dB.
The performance error values are summarized in Table \ref{scalable_table}.
As the microgrid size is scaled from 4 DGs to 10 DGs, we observe a reduction in performance (marked by higher error magnitudes) irrespective of the performance metric.
Further, this reduction becomes more pronounced in the presence of added synthetic noise levels. This can be considered a limitation of the proposed abnormality estimation mechanism.
To understand the framework's practical feasibility, we also verify the abnormality estimation model's robustness and real-time decision-making capabilities in the presence of FDI and MITM attack vectors (manipulations initiated at $t = 7$ s) within the MATLAB-based 10-DG microgrid.
The MATLAB-based system does not involve any noise addition during the real-time evaluation of the ANN-based decision-making framework.
The FDI attack is injected through ($N-1$) transmitters simultaneously.
The MITM attack is introduced in a coordinated manner from the repeaters between DGs 2-3, 5-6, 7-8, 8-9, and 9-10.
As shown in Fig. \ref{fig:scalable}, the proposed framework is resilient against both FDI and MITM attacks in the microgrid.
A comparative analysis of the proposed method's operational details and performance with other state-of-the-art techniques is shown in Table \ref{comparison}. The superiority of the proposed method can be established in terms of resiliency to attacks, mitigation time, and training requirements.

\section{Conclusion}
This paper presented a physics-guided deep ANN model to estimate the possibility of abnormal secondary control operations due to communication-level cyberattacks.
If an attack is identified, a flag is raised (introducing a hold of the last measured stable states) and the ANN checks the set of pre-defined spanning trees to find one that can achieve resilience. Then this topology is enforced mitigating the attack which finally leads to the achievement of nominal operations within the microgrid.
Our results showed that the proposed method is resilient to both transmitter-level FDIs and repeater-level MITM attacks. Further, the performance of the proposed framework also showed robustness irrespective of varying microgrid sizes. However, performance degradation was observed in the presence of noise in higher microgrid sizes. This can be considered a limitation. Future work in this direction will seek to incorporate noise-resilience in the developed framework.



\bibliographystyle{IEEEtran}
\bibliography{biblio}

\end{document}